\documentclass{elsarticle}

\pdfoutput=1

\usepackage{geometry}

\usepackage[T1]{fontenc}
\usepackage[utf8]{inputenc}
\usepackage{lmodern}
\usepackage{relsize}

\usepackage{xcolor}

\usepackage{graphicx}

\usepackage{verbatim}
\usepackage{listings}
\lstdefinestyle{customcpp}{
  breaklines=true,
  xleftmargin=\parindent,
  language=C++,
  showstringspaces=false,
  basicstyle=\small\ttfamily,
  keywordstyle=\bfseries\color{green!40!black},
  commentstyle=\itshape\color{purple!40!black},
  identifierstyle=\color{blue},
  stringstyle=\color{orange},
  tabsize=2,
  escapeinside={/@}{@/},
  keepspaces=true,
  captionpos=b,
  deletekeywords={for,private},
  morekeywords={[2]omp,parallel,sections,single,critical,ordered,barrier,section,master,dummy},
  morekeywords={[2]directive,firstprivate,private,shared,schedule,guided,declare,reduction,initializer,omp_priv,dummy},
  morekeywords={[2]omp_out,omp_in,for,num_threads,flush,atomic,nowait,copyprivate,dummy},
  morekeywords={[2]read,write,update},
  keywordstyle={[2]\bfseries\color{blue!40!black}},
  numbers=left,
}
\usepackage{xspace}
\newcommand\lsttable[2][]{\lstinline[#1]{#2}\xspace}

\usepackage{amsmath}
\usepackage{amsthm}
\usepackage{amssymb}

\usepackage{hyperref}
\usepackage[capitalize]{cleveref}
\newcommand{\cpp}{C\nolinebreak[4]\hspace{-.05em}\raisebox{.4ex}{\relsize{-3}{\textbf{++}}}}
\newcommand{\aeq}{\kern.35em\text{\small+}\kern-.35em=}

\date{}

\begin{document}

\begin{frontmatter}
    \title{Hybrid parallel discrete adjoints in SU2}

    \author[scicomp]{Johannes Bl{\"u}hdorn\corref{cor}}
    \ead{johannes.bluehdorn@scicomp.uni-kl.de}
    \author[su2foundation]{Pedro Gomes}
    \ead{pedro@su2foundation.org}
    \author[scicomp]{Max Aehle}
    \ead{max.aehle@scicomp.uni-kl.de}
    \author[scicomp]{Nicolas R.~Gauger}
    \ead{nicolas.gauger@scicomp.uni-kl.de}

    \affiliation[scicomp]{
        organization={Chair for Scientific Computing, University of Kaiserslautern-Landau (RPTU)},
        addressline={Paul-Ehrlich-Straße 34/36},
        city={Kaiserslautern},
        postcode={67663},
        country={Germany}
    }

    \affiliation[su2foundation]{
        organization={SU2 Foundation},
        addressline={Kluyverweg 1},
        city={Delft},
        postcode={2629 HS},
        country={The Netherlands}
    }

    \cortext[cor]{Corresponding author.}

    \begin{keyword}
        Automatic differentiation \sep High-performance computing \sep Discrete adjoint \sep OpenMP \sep OpDiLib \sep SU2
    \end{keyword}

    \begin{abstract}
        The open-source multiphysics suite SU2 features discrete adjoints by means of operator overloading automatic differentiation (AD). While both primal and discrete adjoint solvers support MPI parallelism, hybrid parallelism using both MPI and OpenMP has only been introduced for the primal solvers so far. In this work, we enable hybrid parallel discrete adjoint solvers. Coupling SU2 with OpDiLib, an add-on for operator overloading AD tools that extends AD to OpenMP parallelism, marks a key step in this endeavour. We identify the affected parts of SU2's advanced AD workflow and discuss the required changes and their tradeoffs. Detailed performance studies compare MPI parallel and hybrid parallel discrete adjoints in terms of memory and runtime and unveil key performance characteristics. We showcase the effectiveness of performance optimizations and highlight perspectives for future improvements. At the same time, this study demonstrates the applicability of OpDiLib in a large code base and its scalability on large test cases, providing valuable insights for future applications both within and beyond SU2.
    \end{abstract}

\end{frontmatter}

\section{Introduction}
\label{section:introduction}

Discrete adjoints are an established methodology to obtain derivatives of simulation codes, e.\,g.\ computational fluid dynamic (CFD) solvers, with numerous applications such as design optimization \cite{VerstraeteMM2017,HeMMM2018}, uncertainty quantification \cite{HuK2019}, parameter identification \cite{AurouxG2017,LaussOSN2018}, or adaptive mesh refinement \cite{KastF2013}. This involves differentiating the discretized problem's solution procedure, where automatic differentiation (AD) has proven its value as a tool that provides accurate, consistent, and maintainable derivatives with bounded performance costs \cite{GriewankW2008,Naumann2011}. Examples for applications of AD to CFD codes include \cite{BischofEtAl1992,ShermanEtAl1996,MuellerC2005,BischofEtAl2006,TowaraN2013,TowaraSN2015}.

With computational resources becoming more powerful and easily available, it is nowadays common to apply simulations to large-scale problems in the order of 100 million degrees of freedom distributed across large numbers of compute nodes. To retain scaling efficiency, simulation codes typically need to be adapted to characteristics of the HPC architecture. With increasing numbers of CPU cores per node, it becomes worthwhile to transition to a hierarchical parallelization approach \cite{Rabenseifner2009} and complement distributed memory parallelism (often implemented via MPI \cite{MPI2023}) with shared memory parallelism, for example via OpenMP \cite{OpenMP52}. For discrete adjoints, this poses the challenge to adapt the AD approaches accordingly. For example, \cite{HueckelheimHSM2019} discusses a strategy to reintroduce OpenMP parallelism to a serial discrete adjoint code obtained with the source transformation AD tool Tapenade \cite{HascoetP2013} from an OpenMP parallel primal code, with applications to a flow solver. Tapenade was subsequently extended to handle OpenMP parallel loops directly \cite{HueckelheimH2022}.

\begin{figure}
    \begin{center}
        \includegraphics{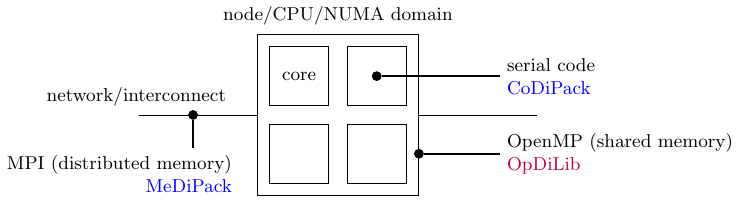}
        \caption{OpenMP-MPI hybrid parallel execution of SU2. The hybrid parallelism is reflected in the AD tools that provide the derivatives. CoDiPack and MeDiPack are already applied to SU2. In the course of this work, we apply OpDiLib for the differentiation of OpenMP.}
        \label{figure:hybrid_parallel_ad}
    \end{center}
\end{figure}

SU2\footnote{\url{https://su2code.github.io/} and \url{https://github.com/su2code} (both visited on July 16th, 2024)} is an open-source code for multiphysics simulations \cite{EconomonPCTA2016}, with applications, e.\,g., in aeroelasticity \cite{GomesP2022} or conjugate heat transfer \cite{BurghardtEtAl2021}. SU2 features discrete adjoints \cite{AlbringSG2015, AlbringSG2016}, using derivatives provided by the operator overloading AD tool CoDiPack\footnote{\url{https://scicomp.rptu.de/software/codi/} and \url{https://github.com/scicompkl/codipack} (both visited on July 16th, 2024)} \cite{SagebaumAG2019}. The MeDiPack add-on\footnote{\url{https://scicomp.rptu.de/software/medi/} and \url{https://github.com/scicompkl/medipack} (both visited on July 16th, 2024)} contributes capabilities to differentiate MPI-parallel codes and enables MPI-parallel discrete adjoints \cite{SagebaumEtAl2023}. After explorative performance studies and prototype developments towards HPC improvements in the context of SU2 and initial improvements of the SU2 code \cite{EconomonEtAl2015,EconomonEtAl2016}, a recent HPC modernization improves the scalability of SU2's primal solvers in multiple ways \cite{GomesEP2021}. For example, changes to the storage layout of solution variables improve the memory locality, and SIMD types ensure that SU2 benefits from the vectorization hardware of today's processors. Furthermore, a new layer of OpenMP parallelism complements the existing MPI parallelism in SU2. It enables hybrid parallel execution as displayed in \cref{figure:hybrid_parallel_ad}. Within a single node (or CPU, or NUMA domain), distributed memory parallelism can be replaced by shared memory parallelism, with perspectives for performance improvements as discussed in the literature \cite{Rabenseifner2009}. For example, fewer MPI processes lead to a reduced communication overhead. At the same time, there is less data duplication across processes, and fewer communication buffers are required, which can reduce per-node memory usage.

In this work, we extend the hybrid parallelism in SU2 to the discrete adjoint solvers, thereby making these advantages available to them as well. As a key step towards enabling hybrid parallel discrete adjoints, we apply OpDiLib\footnote{\url{https://scicomp.rptu.de/software/opdi/} and \url{https://github.com/scicompkl/opdilib} (both visited on July 16th, 2024)}, a novel add-on for operator overloading AD tools that enables the differentiation of OpenMP parallel codes \cite{BluehdornSG2023}. The resulting hybrid differentiation approach is visualized in \cref{figure:hybrid_parallel_ad}. As such, this work also demonstrates the applicability of OpDiLib to a large code base and its performance on large-scale test cases. While our focus is on single-zone problems and SU2's Reynolds-averaged Navier-Stokes (RANS) solver \cite{EconomonPCTA2016}, our insights and improvements are not specific to this setup, so that our work can also serve as a reference both for extending hybrid parallel discrete adjoints to other solvers within SU2 and for applications of OpDiLib to other software packages and AD workflows.

The outline of this paper is as follows. \cref{sec:background} summarizes the background of our work. We provide a brief introduction to discrete adjoints and AD in \cref{sec:discrete_adjoints_and_ad}. \cref{sec:ad_workflow} highlights the components of SU2's AD workflow, as all of these need to be regarded for the transition to hybrid parallel discrete adjoints that we address in detail in \cref{sec:methods_implementation}. \cref{sec:enabling_hybrid_parallel_discrete_adjoints_in_su2} focuses on the transition from MPI-parallel to hybrid parallel discrete adjoints. We describe optimizations of the hybrid parallel discrete adjoint performance in \cref{sec:performance_optimizations}. In \cref{sec:benchmarking}, we evaluate the performance characteristics of our implementation by means of various performance studies. In \cref{sec:single_socket}, we assess the single-socket performance of hybrid parallel discrete adjoints in detail and showcase the effectiveness of the optimizations. We study the performance impact of loop parallelization strategies with respect to hybrid parallel discrete adjoints in \cref{sec:loop_parallelization}. \cref{sec:large_scale} focuses on the performance at large scale. Throughout, we discuss the results and identify promising targets for further performance improvements. We conclude our work in \cref{sec:conclusion}.

\section{Background}
\label{sec:background}

\subsection{Discrete adjoints and automatic differentiation}
\label{sec:discrete_adjoints_and_ad}

Discrete adjoints in SU2 are based on a fixed-point formulation of the primal equations and an interpretation of the primal solution process as a fixed-point iteration \cite{AlbringSG2015, AlbringSG2016}. We summarize the core ideas. Consider the implicit relation
\begin{equation}
    \label{eq:primal_fixed_point}
    U=G(U,X)
\end{equation}
with state $U$, parameters of interest $X$ (e.\,g., mesh coordinates), and an operator $G$ that represents one step of the primal solver. The primal solver in SU2 iterates \eqref{eq:primal_fixed_point} as $U_{n+1}=G(U_n,X)$ until convergence to the primal solution $U^\ast$.

Given an objective function $J$ depending on $U$ and $X$, a Lagrange multiplier approach for the minimization of $J$ with respect to $X$ subject to the constraint \eqref{eq:primal_fixed_point} yields discrete adjoint equations
\begin{align}
    \label{eq:adjoint_fixed_point}
    \bar{U}&=\frac{\partial }{\partial U}J^\mathrm{T}(U^\ast,X)+\frac{\partial }{\partial U}G^\mathrm{T}(U^\ast,X)\bar{U},\\
    \label{eq:adjoint_sensitivity}
    \bar{X}&=\frac{\partial }{\partial X}J^\mathrm{T}(U^\ast,X)+\frac{\partial }{\partial X}G^\mathrm{T}(U^\ast,X)\bar{U},
\end{align}
where the adjoint solution $\bar{U}$ is a suitably chosen Lagrange multiplier and $\bar{X}$ stands for the gradient of the objective function with respect to the parameters $X$ \cite{AlbringSG2015, AlbringSG2016}. Alternatively, \cref{eq:adjoint_fixed_point,eq:adjoint_sensitivity} are obtained by formulating a reverse accumulation procedure for the differentiation of the fixed point \eqref{eq:primal_fixed_point} according to \cite{Christianson1994}. \eqref{eq:adjoint_fixed_point} corresponds to a fixed point iteration for the computation of $\bar{U}$. It is iterated until convergence to $\bar{U}^\ast$, which is then used to compute $\bar{X}$ according to \eqref{eq:adjoint_sensitivity}. Under suitable assumptions, the adjoint fixed-point iteration inherits the convergence properties of the primal fixed-point iteration \cite{Christianson1994}.

The reverse mode of AD is capable of computing the transposed-matrix vector products in \cref{eq:adjoint_fixed_point,eq:adjoint_sensitivity} without assembling any of the involved matrices explicitly. With this capability, the linear system of equations (LSE)
\begin{equation}
\label{eq:adjoint_linear_system}
\left(\frac{\partial }{\partial U}G^\mathrm{T}(U^\ast,X)-I\right)\bar{U}=-\frac{\partial }{\partial U}J^\mathrm{T}(U^\ast,X)
\end{equation}
for $\bar{U}$ implied by \eqref{eq:adjoint_fixed_point} can also be solved by other matrix-free approaches such as Krylov methods \cite{GomesP2022_2}. Besides the fixed-point approach, in SU2 it is possible to use GMRES to solve \eqref{eq:adjoint_linear_system} \cite{GomesP2022}.

We briefly introduce the reverse mode of AD. See, e.\,g., \cite{GriewankW2008,Naumann2011} for comprehensive introductions to AD. We view a computer program $F$ after evaluation of the control flow as a computational graph that translates inputs $x$ via intermediate variables $v$ to outputs $y$. Nodes correspond to variables and edges represent the data flow. Each node is computed from its predecessors by means of an elementary mathematical operation $\phi$ with known partial derivatives as
\begin{equation}
    \label{eq:primal_statement}
    w=\phi(u_1,u_2,\ldots)
\end{equation}
where $w$ is an intermediate variable or an output and the $u_i$ are intermediate variables or inputs. For each variable $v$, we introduce a corresponding adjoint variable $\bar{v}=\frac{\mathrm{d}y}{\mathrm{d}v}^\mathrm{T}\bar{y}$ that represents the components of the derivative of $y$ viewed as a function of $v$ weighted linearly according to user-provided seeds $\bar{y}$. Adjoint variables can be computed in an incremental fashion by reversing the data flow in the computational graph. Starting at the adjoint variables $\bar{y}$ associated with the output variables, each statement \eqref{eq:primal_statement} is complemented by adjoint updates
\begin{equation}
    \label{eq:adjoint_updates}
    \bar{u}_i\aeq\frac{\partial}{\partial u_i}\phi(u_1,u_2,\ldots)\bar{w}
\end{equation}
for all $i$. A proper reset of $\bar{w}$ that regards both adjoint variable reuse and the special case of the same adjoint variable appearing on both sides of \eqref{eq:adjoint_updates} is achieved as follows: create a temporary copy of $\bar{w}$, reset $\bar{w}=0$, and use the temporary copy to perform the updates \eqref{eq:adjoint_updates} \cite{GriewankW2008}. On the level of the whole program, this procedure corresponds to computing
\begin{equation}
    \bar{x}=\frac{\mathrm{d}}{\mathrm{d} x}F(x)^\mathrm{T}\bar{y},
\end{equation}
that is, a transposed-Jacobian vector product such as needed in \cref{eq:adjoint_fixed_point,eq:adjoint_sensitivity}.

\begin{figure}
\begin{lstlisting}[caption={Schematic computation type for operator overloading AD.}, label={listing:su2double}]
struct su2double {
    double value;
    int identifier;
};
\end{lstlisting}
\end{figure}
The reverse mode of AD as described above is provided to SU2 by the operator overloading AD tool CoDiPack \cite{AlbringSG2015,SagebaumAG2019}. As the operator overloading approach relies on exchanging the floating point computation type, SU2 provides a global \lsttable{typedef su2double} and uses it consistently as the floating point computation type throughout the code. Non-AD builds define it as \lsttable{double}, whereas AD builds set it to a suitable CoDiPack type. The latter can be thought of as a structured type as depicted in \cref{listing:su2double} for which, amongst others, the common mathematical operators such as $+$ or $\cdot$, math library functions such as $\sin$ or $\exp$, and assignment operators such as $=$ are overloaded. The overloads perform the usual primal computations and store their results in the \lsttable{value} members of \lsttable{su2double} variables. In addition, the overloads record the computational graph in a stack-like structure commonly referred to as tape. As a part of this, they manage (assign) the \mbox{\lsttable{identifier}} members of \lsttable{su2double} variables \cite{SagebaumBG2020}. The computational graph is recorded in terms of these identifiers. To perform the incremental updates \eqref{eq:adjoint_updates}, the reverse evaluation of the computational graph uses the identifiers to address into CoDiPack's internal vector of adjoint variables (that contains the $\bar{u}_i$ and $\bar{w}$ in \eqref{eq:adjoint_updates}, not to be confused with $\bar{U}$ and $\bar{X}$ in the discrete adjoint equations \eqref{eq:adjoint_fixed_point} and \eqref{eq:adjoint_sensitivity}).

\subsection{Automatic differentiation workflow in SU2}
\label{sec:ad_workflow}

We describe the AD workflow in SU2 with a focus on the discrete adjoint approach given by \eqref{eq:adjoint_fixed_point} and \eqref{eq:adjoint_sensitivity}. First, the primal equation \eqref{eq:primal_fixed_point} is iterated until convergence to $U^\ast$ with a non-AD build of SU2 and subsequently loaded in an AD build of SU2. \eqref{eq:adjoint_fixed_point} assumes a clean separation of all computations into $G$ and $J$, which can be difficult to achieve in practice. For convenience, a primary recording therefore captures the computational graph of the joint evaluation of $G$ and $J$
\begin{equation}
    \label{eq:primary_recording}
    U^\ast\mapsto\begin{pmatrix}G(U^\ast,X)\\J(G(U^\ast,X),X)\end{pmatrix},
\end{equation}
that is, while the actually recorded computation is $J(G(U^\ast,X),X)$ with $U^\ast$ registered as input, the result of $G(U^\ast,X)$ is also treated as an output for which we can set adjoint seeds. Using the current iterate $\bar{U}_i$ as said seeds for the adjoint variables associated with the output of $G$, and 1.0 as the seed for the adjoint variable associated with the output of $J$, an evaluation of this recording corresponds to the iteration
\begin{equation}
    \label{eq:adjoint_fixed_point_adapted}
    \bar{U}_{i+1} =\frac{\partial }{\partial U}\left(J(G(U^\ast,X),X)\right)^\mathrm{T}\cdot 1.0+\frac{\partial }{\partial U}G^\mathrm{T}(U^\ast,X)\bar{U}_i,
\end{equation}
which is of the form \eqref{eq:adjoint_fixed_point} with the modified objective $\tilde{J}(U,X)=J(G(U,X),X)$ (note the additional application of $G$). The updated values $\bar{U}_{i+1}$ are found in the adjoint values associated with the input $U^\ast$ to $G$. SU2 stores the iterate $\bar{U}_i$. For each iteration of \eqref{eq:adjoint_fixed_point_adapted}, SU2 copies the current iterate $\bar{U}_i$ into CoDiPack's internal vector of adjoint variables as seeds, evaluates the tape, and extracts the updated values $\bar{U}_{i+1}$. CoDiPack's vector of adjoint variables is reset to zero in between iterations. After the fixed-point iteration \eqref{eq:adjoint_fixed_point_adapted} has converged to $\bar{U}^\ast$, a secondary recording
\begin{equation*}
    X\mapsto\begin{pmatrix}G(U^\ast,X)\\J(G(U^\ast,X),X)\end{pmatrix},
\end{equation*}
is performed in the same spirit as \eqref{eq:primary_recording}, with the parameters $X$ as inputs, and a single evaluation with respective seeds $\bar{U}^\ast$ and 1.0 corresponds to the realization of \eqref{eq:adjoint_sensitivity} with the same modified objective $\tilde{J}$ as before. In between these two main recordings, SU2 performs an additional evaluation of $J(G(U^\ast,X),X)$ without recording. As a side effect, this passive evaluation resets the \lsttable{identifier} members of all intermediate and output variables to zero, thus resetting the association with adjoint variables. Note that the identifiers of input variables are reset separately, as there are no statements in the course of $J(G(U^\ast,X),X)$ that overwrite them. Note also that $\bar{X}$ admits the interpretation as adjoint variables
\begin{equation*}
    \bar{X}=\frac{\mathrm{d}}{\mathrm{d}X}\tilde{J}(U^\ast(X),X)^\mathrm{T},
\end{equation*}
in the spirit of the reverse mode of AD, that is, the gradient of the objective function with respect to the parameters.

Recording tapes incurs substantial memory costs. Besides improving the accuracy of the derivative computation, a key benefit of the iterative approach described above is hence that it is sufficient to record the computational graph of a single iteration of $G$, as opposed to recording the whole primal iterative process in a black-box fashion. SU2 makes use of further advanced AD techniques to improve the derivative's accuracy and to reduce the memory consumed by recordings \cite{AlbringSG2016}. $G$ commonly involves the approximate solution of LSEs as an iterative subroutine. The analytic derivative of solving an LSE is known \cite{Giles2008} and involves solving an LSE with the transposed matrix, which can be done explicitly using the same iterative LSE solvers \cite{AlbringSG2016}. In terms of AD, LSE solution procedures are treated as externally differentiated functions, or external functions in short \cite{Kowarz2008}. Finally, SU2 makes heavy use of local preaccumulation, that is, contracting computational subgraphs \cite{GriewankW2008}. If a subgraph's number of input and output variables is small compared to its number of intermediate variables, it is more memory-efficient to compute and store the full Jacobian of this subgraph instead of storing the Jacobians for the individual nodes. The Jacobians are assembled by tape recordings and evaluations for subgraphs throughout the recording of the overall tape for $G$. Noting that smaller tapes require less time to evaluate and given that a single iteration of $G$ is recorded once and evaluated multiple times in the course of \eqref{eq:adjoint_fixed_point}, the runtime effort of preaccumulating the Jacobians is compensated for by multiple tape evaluations at reduced runtime cost.

\section{Methods and implementation}
\label{sec:methods_implementation}

\subsection{Enabling hybrid parallel discrete adjoints in SU2}
\label{sec:enabling_hybrid_parallel_discrete_adjoints_in_su2}

Our goal of enabling hybrid parallel discrete adjoints in SU2 affects all parts of the AD workflow discussed in \cref{sec:ad_workflow}. As a key step, we extend the AD capabilities to handle the OpenMP parallel parts in $G$ and $J$, including the advanced AD concepts applied in SU2. We identify and remedy code patterns that do not agree well with hybrid parallel AD, and discuss the involved tradeoffs. We also parallelize the AD workflow itself, in particular the management parts. We properly integrate the new features into the existing code structures and seek to preserve key properties of SU2 such as extensibility and maintainability. While we describe changes made to SU2, many of the observations are general. They hold likewise for enabling hybrid parallel discrete adjoints of other solvers within SU2, and can be transferred to applications of OpDiLib beyond SU2.

\lstset{basicstyle=\footnotesize\ttfamily}
\begin{figure}
\begin{minipage}{0.48\textwidth}
\begin{lstlisting}
SU2_OMP_PARALLEL
{
  SU2_OMP_FOR_(nowait)
  for (int i = 0; i < n; ++i)
  { ... }
  ...
  SU2_OMP_BARRIER
  ...
}
\end{lstlisting}
\end{minipage}
\begin{minipage}{0.48\textwidth}
\begin{lstlisting}[mathescape]
#pragma omp parallel
{
$~~\,$#pragma omp for nowait
  for (int i = 0; i < n; ++i)
  { ... }
  ...
$~~\,$#pragma omp barrier
  ...
}
\end{lstlisting}
\end{minipage}

\hrule

\begin{minipage}{0.48\textwidth}
\begin{lstlisting}
SU2_OMP_PARALLEL
{
  SU2_OMP_FOR_(SU2_NOWAIT)
  for (int i = 0; i < n; ++i)
  { ... }
  END_SU2_OMP_FOR
  ...
  SU2_OMP_BARRIER
  ...
}
END_SU2_OMP_PARALLEL
\end{lstlisting}
\end{minipage}
\begin{minipage}{0.48\textwidth}
\begin{lstlisting}
OPDI_PARALLEL()
{
  OPDI_FOR(OPDI_NOWAIT)
  for (int i = 0; i < n; ++i)
  { ... }
  OPDI_END_FOR
  ...
  OPDI_BARRIER()
  ...
}
OPDI_END_PARALLEL
\end{lstlisting}
\end{minipage}
\begin{lstlisting}[caption={OpenMP parallel code written according to SU2's macro interface (top left) and the corresponding final macro expansion (top right). Hybrid AD compatible code written according to the extended macro interface (bottom left) and the corresponding intermediate macro expansion to OpDiLib macros (bottom right).}, label={listing:su2_macro_expansion_opdi}]
\end{lstlisting}
\end{figure}
\lstset{basicstyle=\small\ttfamily}

\paragraph{Coupling SU2, CoDiPack, and OpDiLib}

CoDiPack has already been applied to SU2 for the purpose of serial and, together with MeDiPack, MPI-parallel discrete adjoints \cite{AlbringSG2016}. SU2 already supports the exchange of the floating point computation type by means of a global \lstinline{typedef su2double}. An AD tool wrapper, implemented by functions in a namespace \lsttable{AD}, abstracts the interaction with CoDiPack. Furthermore, an OpenMP abstraction layer provides \lsttable{SU2_OMP_*} macros that replace OpenMP \lsttable{#pragma} directives throughout the SU2 code, an example is given in the top part of \cref{listing:su2_macro_expansion_opdi}. We build on these facilities for enabling hybrid parallel discrete adjoints.

We update SU2 to use version 2.1 of  CoDiPack, which introduces stable support for OpDiLib. Following the OpenMP parallelism in $G$ and $J$, OpDiLib allows us to record their computational graphs in an OpenMP parallel fashion and employ a corresponding parallelism for the evaluation of the derivative. As detailed in \cite{BluehdornSG2023}, OpDiLib features two strategies (backends) for the detection of OpenMP constructs so that they can be augmented and differentiated accordingly: a fully automatic approach that leverages the OpenMP Tools Interface (OMPT) \cite{EichenbergerEtAl2013}, and a semi-automatic approach that, in particular, requires using \lsttable{OPDI_*} macros instead of OpenMP \lsttable{#pragma} directives, while preserving compatibility with the OMPT backend for compilers that support OMPT. We pursue the semi-automatic approach so that hybrid parallel discrete adjoints in SU2 can make use of OMPT but do not depend on it. We revise the \lsttable{SU2_OMP_*} macros so that they admit expansion to a corresponding \lsttable{OPDI_*} macro. To match OpDiLib's \lsttable{OPDI_NOWAIT} replacement clause for \lsttable{nowait}, we introduce \lsttable{SU2_NOWAIT}. As OpDiLib requires corresponding \lsttable{OPDI_END_*} macros after each OpenMP directive that is followed by a structured block, we also introduce corresponding \lsttable{END_SU2_OMP_*} macros that expand to the \lsttable{OPDI_END_*} macros and apply them throughout the SU2 code. The bottom part of \cref{listing:su2_macro_expansion_opdi} displays an example for hybrid AD compatible OpenMP parallel code in SU2 and the corresponding intermediate expansion to OpDiLib macros. When building SU2 with hybrid AD support, the OpDiLib backend is chosen automatically according to the detected compiler support for OMPT. Optionally, the user can override this with the \texttt{opdi-backend} build option.

We perform the initialization and finalization of OpDiLib in the new \lsttable{omp_initialize()} and \mbox{\lsttable{omp_finalize()}} functions in the OpenMP abstraction layer. The AD tool wrapper becomes aware of OpDiLib's internal states, which complement tapes and evolve alongside the recordings. We match exports and recoveries of tape positions by exports and a recoveries of OpDiLib's state, respectively. We account for OpDiLib-specific preparations and cleanups by notifying OpDiLib prior to and after tape evaluations. As tapes are thread-local objects that change at runtime, we resolve all tape references dynamically via \lsttable{AD::getTape()}.

\paragraph{Compatibility with reuse identifier management schemes} For serial and MPI-parallel discrete adjoints, SU2 defines \lsttable{su2double} to be CoDiPack's type \lstinline{codi::RealReverse}. For hybrid parallel discrete adjoints, we rely on the OpenMP-compatible CoDiPack type \lsttable{codi::RealReverseIndexOpenMP} (more precisely, an instance of the template \lsttable{codi::RealReverseIndexOpenMPGen} that ensures non-atomic evaluations for serial parts of the code, see \cref{sec:performance_optimizations}). An important difference between these types is the underlying strategy for managing identifiers. \lsttable{codi::RealReverse} features a linear management scheme, where identifiers are assigned in an incremental fashion and each identifier is assigned at most once throughout a recording. \lsttable{codi::RealReverseIndexOpenMPGen} relies on an OpenMP-aware reuse management scheme, where each thread maintains a pool of identifiers from which new identifiers are drawn and to which identifiers of destructed or overwritten variables are released. Identifiers can be associated with different variables throughout a recording as long as an identifier is not associated with multiple variables at the same time. An overview of identifier management schemes and their properties is given in \cite{SagebaumBG2020}. We identify and eliminate code patterns in SU2 that are not compatible with reuse management schemes, making the discrete adjoint in SU2 at the same time less dependent on specific AD tool properties. We introduce a \texttt{codi-tape} build option that can be used to test compatibility with reuse management schemes also in a non-hybrid setting.

Uninitialized identifier memory, unless it is zero by chance, contains identifiers that were never created and assigned properly and will eventually pollute the pool of identifiers. We therefore initialize memory of all \lsttable{su2double} variables, preferably by constructor calls, or by memsetting to zero. Particular examples are allocations of vectors and matrices for linear algebra operations as well as MPI buffers.

Identifiers that are not reintroduced to the pool of identifiers lead to dead adjoint memory. We therefore guarantee proper releases of identifiers for all AD variables by destructor calls or passive overwrites, as opposed to, e.\,g., setting identifiers to zero manually. In particular, we self-assign to each input variable its own passive value component, as input variables are not overwritten by SU2's passive evaluation of $J$ and $G$ for clearing identifiers.

The linear management scheme of \lsttable{codi::RealReverse} admits simplified copies: the identifier can be copied as well, without recording the statement on the tape and without notifying the AD tool. Due to the shared identifier, copies behave like references in terms of AD, which allows counterintuitive code patterns. Suppose a variable should become an input of a preaccumulated subgraph. Due to the copied identifier, it would be sufficient to only register a copy of this variable as an input, and moreover, to refer to this variable only by copies throughout the subgraph. The reuse management scheme, on the other hand, assigns different identifiers to copies and cannot support such code patterns. To remedy this, we adapt various accessors to global data to return references. We forbid registration of rvalues as preaccumulation inputs. In general, if references are intended, references should be preferred over copies that might or might not share their identifier.

We remark that the taping strategy for SU2's multiphysics solvers \cite{BurghardtEtAl2021} is not compatible with reuse management schemes yet in the case that more than one zone is used. This is due to the joint recording of contributions from multiple zones on a single tape in an interleaved manner and a partial tape evaluation strategy that may skip parts of the tape. Reused identifiers can be associated with different variables in different parts of the tape, which is reflected in the tape evaluation by setting consumed adjoint variables to zero, like the $\bar{w}=0$ in the context of \eqref{eq:adjoint_updates}. If an evaluation omits parts of the tape, it also omits the corresponding adjoint variable resets, so that contributions to different adjoint variables are mixed at the same memory location.

\paragraph{Hybrid parallel AD of the linear solver} As outlined in \cref{sec:ad_workflow}, $G$ relies on subroutines to solve LSEs, which are differentiated analytically by means of solving an LSE with the transposed matrix. Previously, SU2 provided dedicated routines for the multiplication with the transposed matrix, without actually transposing the matrix. This requires specialization with respect to each preconditioner and is inefficient to parallelize with OpenMP as it involves strided memory access. Therefore, we transpose the matrix in place instead, so that the primal LSE solver routines can be reused, including their parallelism and all preconditioners. The improved memory access pattern and increased flexibility should outweigh the cost of transposing the matrix.

All threads work jointly on the externally differentiated LSE solver. CoDiPack offers an OpenMP-aware external function helper for this use case. We leave it to the master thread, surrounded by appropriate barriers, to register external function inputs and outputs as well as external function data such as the LSE's matrix. All threads manage the activity of their tape, register the external function on their tape, participate in solving the corresponding reverse LSE, and fetch registered data such as the LSE's matrix in a thread-safe manner.

\paragraph{OpenMP-parallel preaccumulations} With OpenMP, AD workflows for local preaccumulation become thread-local AD workflows. Nonetheless, all threads share the same vector of adjoint variables and rely on it for evaluating their local recordings. With this design, there is no support for sharing of \lstinline{su2double} variables between simultaneous preaccumulations as it leads to data races. To give an example, consider two simultaneous AD workflows that register the same variable as input. Both workflows would accumulate derivative contributions in the single adjoint variable associated with the common input variable, yielding the sum of two unrelated Jacobian entries instead of two separate Jacobian entries.

In hybrid AD builds, we disable the preaccumulations for which we cannot guarantee no-sharing. We do this either locally, or, for larger parts of the code, in particular edge loops that are parallelized according to the reduction strategy (see \cref{sec:performance_optimizations}), by means of \lsttable{AD::PausePreaccumulation()} and \lsttable{AD::ResumePreaccumulation()} functions, which we introduce in SU2's AD tool wrapper.

SU2 does not use OpenMP constructs within thread-local preaccumulations. While we did not have to revise the code in this regard, it is important to note that OpenMP constructs in simultaneous thread-local AD workflows are not supported either, especially synchronization across AD workflows would be challenging to handle correctly. For instance, OpDiLib's handling of mutex synchronization requires serially executed preparations prior to parallel tape evaluations, a type of synchronization that patterns such as preaccumulations within items of worksharing loops do not admit.

In follow-up research in \cite{BluehdornG2024}, we study the issue of simultaneous preaccumulations with shared inputs in detail, discuss approaches to re-enable such preaccumulations, and test these approaches in the context of hybrid parallel discrete adjoint computations in SU2.

\paragraph{Hybrid parallel discrete adjoint solvers}

Discrete adjoint driver, iteration, and solver classes implement steps of the reverse accumulation AD workflow summarized in \cref{sec:ad_workflow}. Tape recordings and evaluations in a discrete adjoint solver inherit the underlying primal solver's OpenMP parallelism. In addition, we parallelize the (re)computation of dependent quantities after AD input registration as well as the reverse accumulation AD workflow itself. We introduce parallel loops for resetting input identifiers and for copying derivatives, in particular the extraction and initialization of adjoints in between iterations. This requires lookups in the vector of adjoint variables according to previously remembered identifiers of input and output variables. We revise the storage strategy for these identifiers. Previously, identifiers of input and output variables were stored upon AD input registration and output registration, respectively, in a sequential manner in large arrays, and later retrieved with the assumption that the order of variable accesses is always the same. Instead, we store identifiers alongside solution variables in accordance with the revised storage layout \cite{GomesEP2021}. This does not only clarify the relation between identifiers and variables but is also compatible with OpenMP parallel extraction and recovery of identifiers.

\paragraph{Automated testing and thread sanitizer analysis}

We establish tests for the new hybrid AD features that are executed regularly as part of SU2's CI pipeline, which includes build tests of hybrid AD configurations as well as hybrid AD regression tests. The latter are derived from already established MPI-parallel AD regression tests. They validate the new hybrid parallel discrete adjoint features and help to maintain them as SU2 is developed further. 

We were able to hide many of the changes discussed in the previous paragraphs in SU2's abstraction layers for AD and OpenMP, which is in line with the principle that general SU2 development should not require in-depth familiarity with these topics. There are a two notable exceptions. First, the \lsttable{END_SU2_*} macros throughout the SU2 code are uncommon for OpenMP development in \cpp. To mitigate their maintenance effort, we introduce automatic syntax checks as part of SU2's hybrid AD build tests. These checks detect both missing end macros as well as mismatched pairs of opening and closing macros. Second, treating preaccumulations in the presence of sharing as well as applying the shared reading optimization discussed in \cref{sec:performance_optimizations} require certain awareness of the data access patterns. Assumed or declared no-sharing, if violated, can become a new source of data races as the SU2 code is developed further.

Data races can be difficult to identify and debug, in particular if they only occur during tape evaluations, where the relation to the underlying error in the recorded code is often unclear. We therefore apply the thread sanitizer, a tool for the automatic detection of data races \cite{SerebryanyI2009}. We conduct a comprehensive analysis of SU2 and fix various data races both in the primal and in the discrete adjoint code, including violations of declared no-sharing. In the course of this, we introduce \lsttable{SU2_OMP_SAFE_GLOBAL_ACCESS} macros that abbreviate frequently occurring barrier-master-barrier patterns within parallel regions.

To detect data races as early as possible and to mitigate the maintenance effort of hybrid parallel code, both with and without AD, we extend SU2's CI pipeline to automate the thread sanitizer analysis. We provide thread-sanitizer enabled containers as part of the su2code/Docker-Builds repository\footnote{\url{https://github.com/su2code/Docker-Builds} (visited on July 16th, 2024)}, which are not only the basis for the CI pipeline tests but can also be used to test for data races locally, with notably reduced setup and configuration effort.

\paragraph{Remarks on AD and vectorization}

As illustrated in \cref{listing:su2double}, CoDiPack's types for the reverse mode of AD are structured types (floating point value plus integer identifier). Arrays of structs are not suitable for vectorization, neither automatically nor by means of SU2's explicit SIMD types, so that no speedup for \lsttable{su2double} computations can be expected. SIMD types with \lsttable{su2double} and vector width larger than one even lead to a slowdown for preaccumulations because they fuse multiple independent, scalar AD workflows into a single one. Without specific treatment, the full Jacobian is preaccumulated although it is actually sparse. To avoid this blow-up of the Jacobian while retaining the explicit SIMD types in performance-critical parts of the code, we resort to \lsttable{su2double} SIMD types with a vector width of one in AD builds. As LSE solvers (both primal and adjoint) use intrinsic floating point types, AD builds still benefit from vectorization of the LSE solver.

\subsection{Performance optimizations}
\label{sec:performance_optimizations}

We describe two types of optimizations that help to improve the hybrid parallel discrete adjoint performance. They are not specific to SU2 and can also be applied to other AD workflows that incorporate OpDiLib.

\paragraph{Optimized adjoint vector management}

The final size of CoDiPack's vector of adjoint variables is not known in advance. CoDiPack can check for sufficient size automatically and resize (reallocate) the vector as needed, for example when seeding or prior to tape evaluations. In a shared-memory setting, however, all threads collaborate on a shared vector of adjoint variables. To guarantee mutual exclusion of adjoint vector resizing on the one hand and reading and writing of adjoint variables on the other hand, the vector of adjoint variables is protected by a shared mutex, where the exclusive lock corresponds to resizing and the shared lock corresponds to usage. 

To ensure that locking does not become a performance bottleneck, we reduce the number of locking operations by managing the vector of adjoint variables explicitly. We introduce an \lstinline{AD::ResizeAdjoints()} function in SU2's AD tool wrapper with which we ensure sufficient size prior to multiple subsequent adjoint variable accesses. This avoids serialization due to frequent exclusive locking and saves the cost of frequent reallocation. In addition, we modify the existing adjoint variable access routines \lstinline{AD::GetDerivative(...)} and \lstinline{AD::SetDerivative(...)} to perform neither bounds checking nor implicit resizing. To address the overhead of frequent shared locking, we protect subsequent adjoint variable accesses by a single shared lock by means of new \lstinline{AD::BeginUseAdjoints()} and \lstinline{AD::EndUseAdjoints()} functions, instead of defaulting to implicit shared locks for each individual access.

This affects in particular the management parts of SU2's AD workflow, like the seeding and extraction of adjoint variables between discrete adjoint iterations \eqref{eq:adjoint_fixed_point}. Additionally, we improve the preaccumulation performance by implementing the same improvements in CoDiPack for local preaccumulation workflows.

\paragraph{Shared reading optimization}

Consider two statements in the primal code in the spirit of \eqref{eq:primal_statement} that are executed and recorded in parallel and have a common variable on the right hand side,
\begin{equation*}
    w_1=\phi_1(u,\ldots),\hspace{1cm}w_2=\phi_2(u,\ldots).
\end{equation*}
The corresponding adjoint updates on $\bar{u}$ according to \eqref{eq:adjoint_updates} read
\begin{equation*}
    \bar{u}\aeq\frac{\partial}{\partial u}\phi_1(u,\ldots)\bar{w}_1,\hspace{1cm}\bar{u}\aeq\frac{\partial}{\partial u}\phi_2(u,\ldots)\bar{w}_2.
\end{equation*}
As the primal parallelism is mirrored in the tape evaluation, the adjoint updates are also executed in parallel, which leads to a data race on $\bar{u}$. Atomic increments on adjoint variables resolve these data races, see, e.\,g., \cite{Foerster2014,FoersterNU2011}. Parts of the parallel code where different threads do not share variables on right hand sides (referred to as \emph{exclusive read property} in \cite{Foerster2014}), on the other hand, do not require atomic updates. To leverage developer knowledge about the data access patterns, we introduce new \lsttable{AD::StartNoSharedReading()} and \lsttable{AD::EndNoSharedReading()} functions in SU2's AD tool wrapper and use them to mark such parts of the code. Internally, they change OpDiLib's adjoint access mode to use non-atomic increments, and thus avoid the overhead of atomic updates. We remark that neither serial parts of the recording (anything outside parallel regions) nor preaccumulations use atomics as both follow CoDiPack's evaluation procedure associated with the AD type (as opposed to OpDiLib's evaluation procedure for parallel regions that can switch between atomic and non-atomic updates). This is valid for preaccumulations due to their strict no-sharing requirement. In addition, we declare serial parts within parallel regions that are isolated in a barrier-master-barrier fashion as safe for non-atomic evaluation. The shared reading optimization is enabled by default. We add an \texttt{opdi-shared-read-opt} build option to disable it, e.\,g., to study its impact on the evaluation performance.

Further potential to eliminate atomic updates lies in SU2's strategies for loop parallelization.
Given the discretization (i.\,e.\ mesh) of a computational domain, implementations of the finite volume or finite element methods require loops over the discretization elements (control volumes, elements, edges, points), to apply the numerical schemes used to solve the discrete problem.
The two main aspects of enabling shared-memory parallelism of such loops are adding the required work-sharing directives to the code, and developing strategies to avoid data races that would otherwise result from gather-scatter access patterns. For example, in finite-volume solvers, the flux across a face (or edge) depends on the quantities of two adjacent control volumes, and in turn, this flux contributes to the divergence (also known as residual) of both control volumes.
Of the strategies evaluated in \cite{GomesEP2021}, two were chosen for loops with high arithmetic intensity, namely grid coloring and reductions.
In the former, edges and elements are colored such that data races do not occur when threads loop simultaneously over entities with the same color (synchronization is required once per color).
The reduction strategy is a type of scatter-to-gather loop transformation that consists in first storing the results of a loop over edges (without scattering) and then reducing this intermediate variable by looping over the faces of each control volume.
Coloring reduces the locality of the data, whereas the reduction strategy increases storage requirements and introduces a relatively expensive reduction (approximately the cost of one linear solver iteration) without sacrificing locality or parallelism. The caching performance of coloring is improved by using color groups. Within a group, the no-sharing constraint is lifted, so that all groups of a color can be processed in parallel, but each group must be assigned to a single thread. As a rule of thumb, larger color groups result in better caching, but make it more difficult to obtain a coloring with well-balanced parallel workloads.

With a view on the primal performance, SU2 switches to the reduction strategy on all MPI ranks on which coloring with the specified edge color group size fails or yields an efficiency below 0.875 in terms of load balancing across the threads of the rank \cite{GomesEP2021}. The tradeoffs between coloring and reductions are different for discrete adjoints. As was also observed in \cite{HueckelheimHSM2019} in the context of edge loop parallelization in a flow solver, the coloring approach leads to exclusive read access. We can therefore apply the shared reading optimization. The reduction strategy, on the other hand, does not only have inherent simultaneous read accesses and cannot be optimized in this regard but also results in larger tapes that consume more memory and take longer to evaluate. For these reasons, coloring is preferential for discrete adjoints even at the cost of reducing the edge color group size. To increase the success rate of the coloring algorithm, we allow larger numbers of colors by default. To find a coloring that both performs well in terms of caching and is efficient in terms of load balancing, each MPI rank adapts its edge color group size. We assume a monotonic relationship between edge color group size and efficiency and treat coloring success and coloring efficiency of at least 0.875 as constraints. Starting with the specified edge color group size as an upper bound, each MPI ranks determines the largest admissible edge color group size by means of bisections. This adaptive strategy should result in better caching than resorting to coloring without groups (edge color group size one) and is at the same time more flexible and automatic than manual tuning of the edge color group size. The decisions are supported by the performance study in \cref{sec:loop_parallelization}. Adaptivity is enabled by default and can be toggled with the \verb|EDGE_COLORING_RELAX_DISC_ADJ| SU2 configuration file option.

\section{Benchmarking results and discussion}
\label{sec:benchmarking}

Throughout \cref{sec:enabling_hybrid_parallel_discrete_adjoints_in_su2}, we have changed several aspects of the AD implementation in order to enable hybrid parallel discrete adjoints. To understand the impact of these changes on the overall hybrid AD performance as well as the impact of the optimizations described in \cref{sec:performance_optimizations}, we conduct detailed studies of the single-socket performance on two test cases in \cref{sec:single_socket}. In \cref{sec:loop_parallelization}, we study the impact of the loop parallelization strategies discussed in \cref{sec:performance_optimizations}. Finally, we assess the scalability with a large test case in \cref{sec:large_scale}. In all test cases, we start with converged primal solutions of the RANS equations with the Spalart-Allmaras (SA) turbulence model \cite{SpalartA1992}. The fluid is modelled as an ideal gas. We discretize convective fluxes with the JST scheme \cite{JamesonST1981} and compute spatial gradients with the Green-Gauss approach. See \cite{Blazek2015} for an overview over the governing equations and numerical schemes. We configure discrete adjoints to compute derivatives of the aerodynamic lift with respect to mesh coordinates, which could, for example, be used for shape optimization. Our goal is to assess comparable computational workloads across different parallel setups, and we therefore always execute a fixed number of discrete adjoint iterations. In particular, we do not make statements about discrete adjoint convergence. Unless mentioned otherwise, we parallelize loops with edge coloring, using the default edge color group size of 512 and adaptivity for discrete adjoints.

We execute our performance tests on Skylake nodes of the Elwetritsch cluster at the RPTU. Each node features two Intel Xeon Gold 6126 processors at 2.6\,GHz, and each processor consists of 12 cores arranged in a single NUMA domain. Dynamic frequency scaling is disabled. We always consider average measurements of five consecutive runs, after one discarded warm-up run.

\begin{figure}
    \begin{center}
    \includegraphics[width=0.65\textwidth]{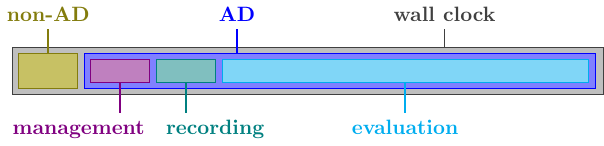}
    \end{center}
    \caption{Decomposition of the discrete adjoint wall clock time.}
    \label{fig:wall_clock_decomposition}
\end{figure}
We decompose the wall clock time of discrete adjoint computations as visualized in \cref{fig:wall_clock_decomposition}. We distinguish between AD-relevant timings on the one hand, that is, all timings that are directly influenced by the changes described in \cref{sec:enabling_hybrid_parallel_discrete_adjoints_in_su2,sec:performance_optimizations}, and non-AD timings on the other hand, like preprocessing of the geometry or file I/O. While non-AD timings might be different between MPI-only and hybrid builds of SU2, this is not influenced by hybrid AD itself and not of primary interest in our comparisons. We further refine the AD-relevant timings into recording, evaluation, and management. Recording encompasses the primary and secondary recording passes as explained in \cref{sec:ad_workflow} as well as passive computations for the purpose of clearing identifiers. Evaluation refers to all tape evaluations in the course of discrete adjoint iterations \eqref{eq:adjoint_fixed_point} or \eqref{eq:adjoint_linear_system}, respectively, as well as the final evaluation with respect to parameters in \eqref{eq:adjoint_sensitivity}. Management encompasses interactions with CoDiPack's vector of adjoint variables, that is, the seeding prior to tape evaluations as well as the extraction after tape evaluations.

Support for hybrid parallel discrete adjoints is available in the upstream version of SU2. We make a few changes to ensure comparable computational workloads across different parallel setups (e.\,g., no special behaviour for the case of a single OpenMP thread, lower tolerances so that iterations do not stop early). We use release builds of SU2, compiled with GCC 11.3, with MPI support (OpenMPI 4.0), with AD enabled, with AVX512, and use single precision for linear algebra. The changes made to SU2 and precise build settings are documented in the \verb|hybrid_parallel_discrete_adjoints| branch\footnote{\url{https://github.com/jblueh/SU2/tree/hybrid_parallel_discrete_adjoints} (visited on July 16th, 2024)}, alongside the test cases that are used in the following\footnote{\url{https://github.com/jblueh/SU2/tree/hybrid_parallel_discrete_adjoints/benchmark} (visited on July 16th, 2024)}.

\subsection{Single-socket performance}
\label{sec:single_socket}

\begin{figure}
\begin{minipage}{0.49\textwidth}
\includegraphics[width=1\textwidth]{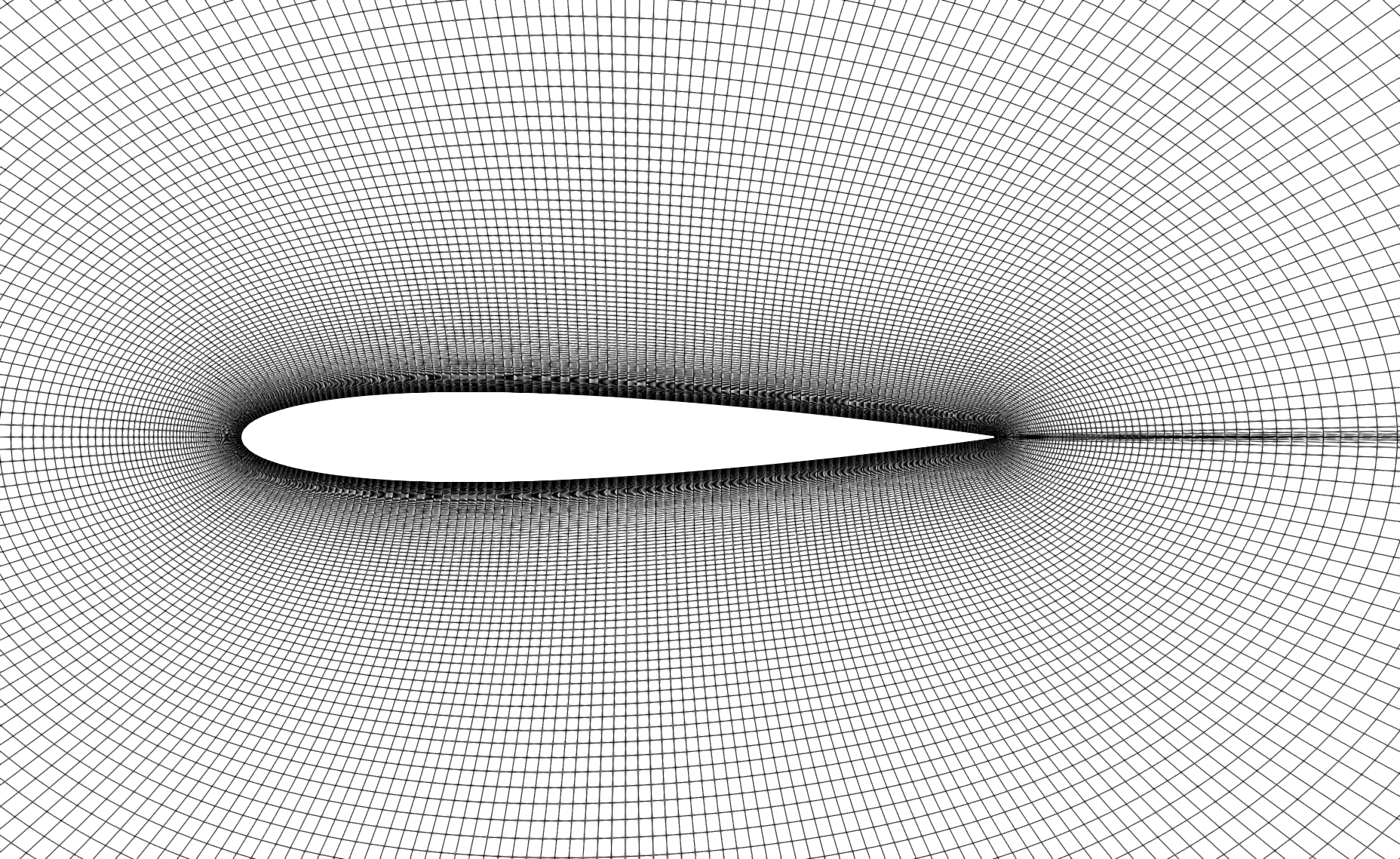}
\end{minipage}
\begin{minipage}{0.49\textwidth}
\includegraphics[width=1\textwidth]{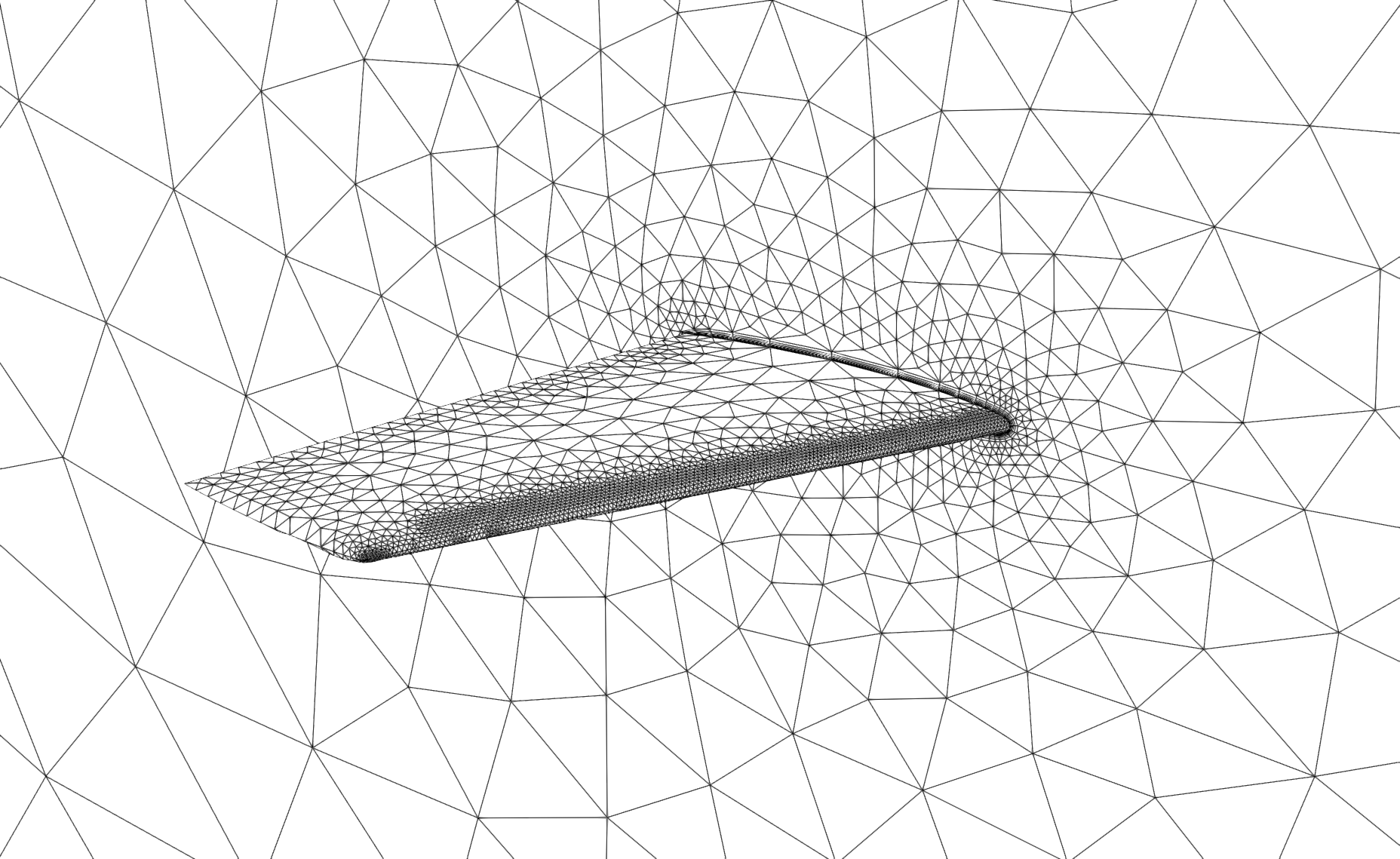}
\end{minipage}
\caption{NACA 0012 mesh (left) and Onera M6 mesh (right) for single-socket performance studies.}
\label{fig:single_socket_meshes}
\end{figure}

We work with two test cases that feature different types of meshes, different flow scenarios, and different approaches for the discrete adjoint computation.
\begin{enumerate}
    \item We consider the flow around a 2D NACA 0012 airfoil at Mach 0.5 and $0^\circ$ angle of attack (AoA), with a Reynolds number (Re) of $20\cdot10^6$. We assume a constant dynamic viscosity of $1.02\cdot10^{-5}\,\text{Pa\,s}$. The computational domain is discretized by means of a structured mesh with 75140 quadrilaterals, as displayed in the left part of \cref{fig:single_socket_meshes}. Given the converged primal flow solution, we perform 300 discrete adjoint iterations in the spirit of \eqref{eq:adjoint_fixed_point} followed by one evaluation of \eqref{eq:adjoint_sensitivity}.
    \item We consider the flow around a 3D Onera M6 wing at Mach 0.8395 and $3.06^\circ$ AoA, $0^\circ$ sideslip angle, $\text{Re}=11.72\cdot 10^6$, and a Reynolds length of $0.64607\,\text{m}$. This resembles flow settings for the Onera M6 wing used in \cite{SchmittC1979}. The dynamic viscosity follows Sutherland's law \cite{Sutherland1893} with parameters for air. An unstructured mesh consisting of 258969 tetrahedrons discretizes the computational domain. \cref{fig:single_socket_meshes} displays the wing surface and parts of the symmetry plane. Starting with the converged primal flow solution, we perform 300 GMRES iterations to solve \cref{eq:adjoint_linear_system}. We restart the Krylov solver after every 50th iteration. Afterwards, we perform one evaluation of \eqref{eq:adjoint_sensitivity}.
\end{enumerate}
We execute the tests on a single cluster node and bind all processes and threads to a single socket.

To understand the differences in the AD-specific timings in detail, we consider various intermediate build configurations of SU2 that illustrate the key steps in the transition to hybrid parallel discrete adjoints. We begin with the classical, MPI-only build of SU2 (\emph{MPI build, linear}). We first exchange the linear identifier management scheme with a reuse scheme (\emph{MPI build, reuse}). Next, we disable preaccumulations that are incompatible with hybrid parallel AD (\emph{MPI build, reuse, hybr.~preacc.}). We then transition to a hybrid build that uses OpDiLib and features all changes described in \cref{sec:enabling_hybrid_parallel_discrete_adjoints_in_su2} (\emph{hybrid build, no opt.}). Finally, we enable the optimizations discussed in \cref{sec:performance_optimizations}, first only optimized adjoint vector management (\emph{hybrid build, opt.~adj.~mgmt.)} and then in addition also the shared reading optimization (\emph{hybrid build, both opt.}).

\begin{figure}
    \includegraphics[width=\textwidth]{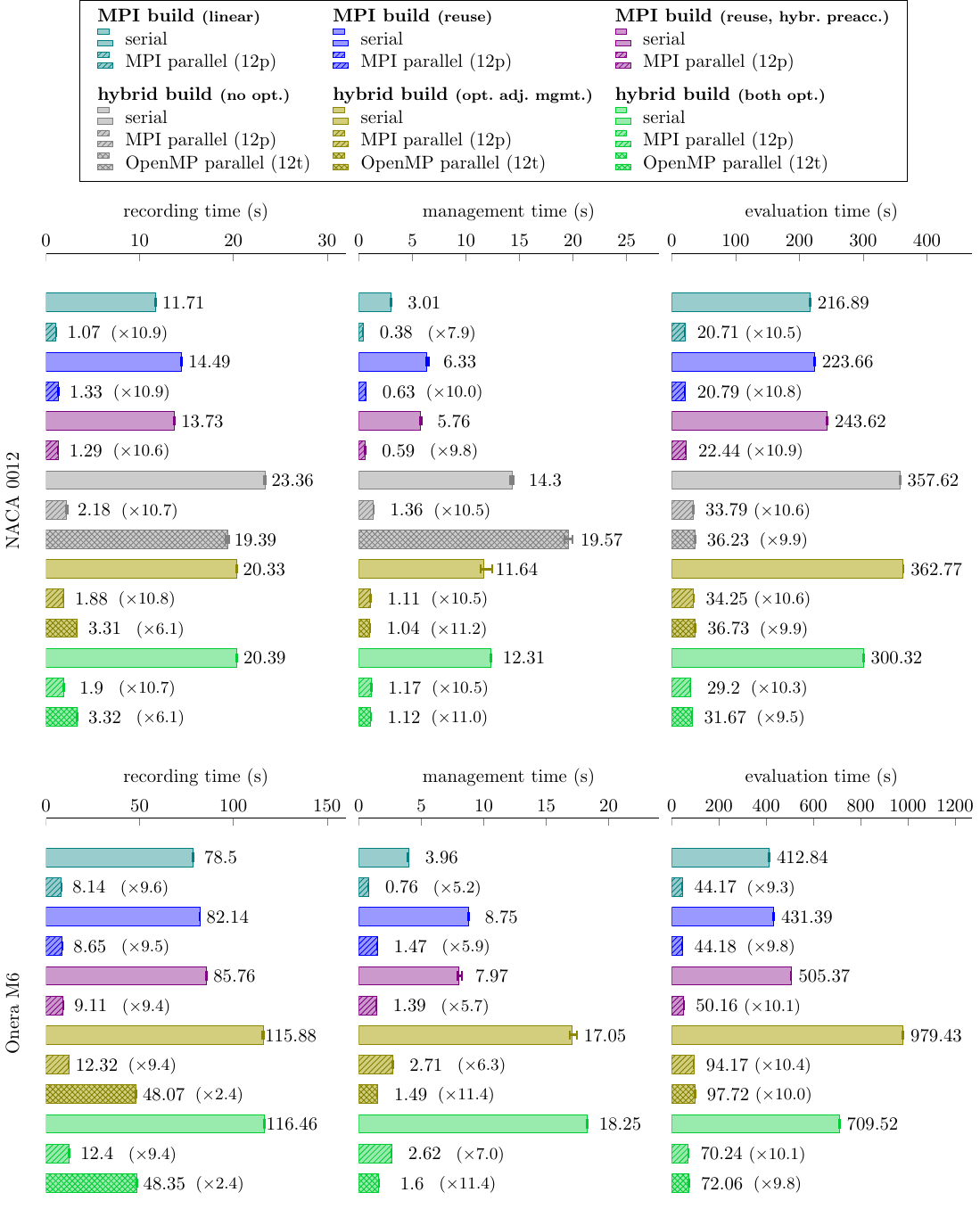}
    \caption{AD-specific performance of the NACA 0012 test case (top) and the Onera M6 test case (bottom). Recording performance (left), management performance (middle), and evaluation performance (right). Serial and parallel timings for various build configurations. Error bars indicate the variation across runs. Speedup factors are relative to the serial run of the respective build.}
    \label{fig:single_socket_performanc}
\end{figure}
The left part of \cref{fig:single_socket_performanc} displays the AD recording performance. The respective serial recording timings, for both the NACA 0012 and the Onera M6 test case, illustrate the key differences between the builds. Switching to a reuse management scheme incurs performance overhead due to increased tape size and more involved identifier management, as explained in detail in \cite{SagebaumBG2020}. Enabling only preaccumulations compatible with hybrid AD reduces the overall number of preaccumulations, which usually reduces the recording time slightly. Switching to a hybrid build requires a thread-safe CoDiPack type, which incurs costs mainly due to tapes accessed via static thread-local pointers \cite{BluehdornSG2023}. Optimizing for adjoint vector management makes the preaccumulations throughout the recording more efficient, also in the serial case. The shared reading optimization has only a minor impact on the recording performance, the observed small overheads could be due to the tracking of code parts suitable for evaluation without atomics.

Despite the differences in the serial performance, we observe consistent MPI speedups in the ranges $\times10.6$ to $\times10.9$ and $\times9.4$ to $\times9.6$ for the NACA 0012 and Onera M6 test cases, respectively. Considering the OpenMP-parallel recording performance, the NACA 0012 timings clearly show that optimizing the adjoint vector management is crucial for the recording to scale, even more so for the management, which is why we do not consider unoptimized hybrid builds in the Onera M6 test case. With optimized adjoint vector management, we observe OpenMP speedups of $\times6.1$ and $\times2.4$ for the two test cases, which are smaller than the respective MPI speedups, in particular for the Onera M6 test case. This is in parts due to preaccumulations, as even with optimized adjoint vector management, they still require frequent locking and reallocation of the vector of adjoint variables. Serial parts of $G$ and $J$ that are not parallelized with OpenMP contribute to the reduced speedup as well, for example the recomputation of volumes of dual control volumes (which are dependent grid quantities).

The management timings show similar trends. Unlike linear identifier management, reuse identifier management does not necessarily assign contiguous identifiers to successively registered inputs and outputs. We suspect that the corresponding less regular memory access pattern causes the performance drop in the transition to reuse identifier management. In the transition to a hybrid build, the management timings are subject to the same overheads as the recording timings. MPI speedups range from $\times7.9$ to $\times10.5$ in the NACA 0012 test case and from $\times 5.2$ to $\times 7.0$ in the Onera M6 test case. With optimized adjoint vector management, we observe OpenMP speedups of $\times11.0$ and $\times11.2$ in the NACA 0012 test case and $\times 11.4$ in the Onera M6 test case.

The right part of \cref{fig:single_socket_performanc} showcases the AD evaluation performance of the two test cases, which both show similar trends. We explain the evolution of the serial performance. Reuse management leads to larger tapes, both in terms of tape entry size and the overall number of tape entries, and correspondingly larger evaluation times \cite{SagebaumBG2020}. Likewise, disabling some of the preaccumulations shifts workload from the recording phase to the evaluation phase. The hybrid build's serial evaluation performance demonstrates the absolute cost of performing atomic updates on adjoint variables. While optimized adjoint vector management does not affect the evaluation performance, the shared reading optimization reduces the overhead by approximately fifty percent in both cases, which underlines the importance of eliminating atomic updates on adjoint variables. We observe consistent MPI speedups of $\times10.3$ to $\times 10.9$ and $\times9.3$ to $\times10.4$ for the NACA 0012 and the Onera M6 test case, respectively, and respective OpenMP speedups in the ranges $\times9.5$ to $\times 9.9$ and $\times9.8$ to $\times10.0$. As the evaluation inherits the parallelism of the recording, serial parts that limit the OpenMP scaling of the recording will limit the scaling of the evaluation as well. In addition, serialization due to atomic updates on adjoint variables could be a limiting factor for the OpenMP speedup. The observed OpenMP speedups are competitive, which is important for the overall performance as the evaluation time scales with the number of discrete adjoint iterations.

\begin{figure}
    \begin{center}
        \includegraphics[width=0.9\textwidth]{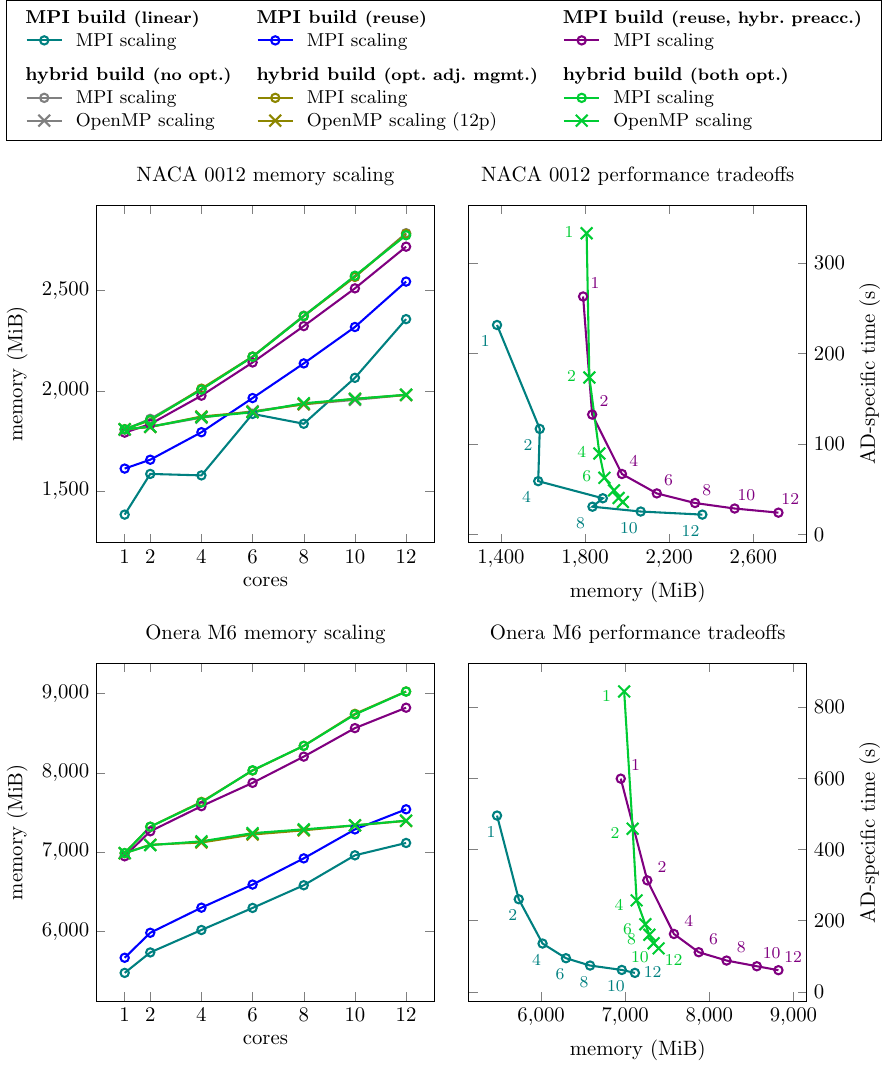}
    \end{center}
    \caption{NACA 0012 performance (top) and Onera M6 performance (bottom). Memory high-water marks depending on the type and degree of parallelism for the five SU2 build configurations (left), and joint memory consumption and evaluation time for selected configurations with varying degrees of parallelism (right).}
    \label{fig:single_socket_memory}
\end{figure}
We investigate the memory consumption of the build configurations with varying degrees of parallelism for both test cases in the left part of \cref{fig:single_socket_memory}. We first explain the differences between the build configurations at serial execution. When we compare the memory consumption of hybrid builds to the memory consumption of the classical MPI-only build with linear management, we see that one part of the difference is due to the switch to reuse management, which requires larger individual tape entries and, as copy operations need to be recorded, at the same time more overall tape entries, see \cite{SagebaumBG2020}, where the benefits of optimized copies for SU2 were observed as well. Another part is due to disabling incompatible preaccumulations, which results in larger tapes and therefore higher memory consumption. While both types of overhead have similar magnitudes for the NACA 0012 test case, the additional memory consumption due to disabled preaccumulations becomes dominant in the Onera M6 test case. The optimizations from \cref {sec:performance_optimizations} do not affect the memory consumption. For each test case, the memory usage increases linearly with an increasing number of MPI processes with similar trends across all build configurations. We attribute this to data duplication across processes, additional halo layers in the partitioning, and larger communication buffers. The memory overhead of increasing the number of OpenMP threads, on the other hand, is significantly smaller, so much that for the NACA 0012 test case, OpenMP parallel execution of hybrid builds requires less per-node memory than MPI parallel execution of the classical build configuration for this test case already at 10-fold parallelism, despite the larger base memory consumption. For the Onera M6 test case, on the other hand, 12-fold parallelism does not suffice to compensate for the memory overhead due to disabled preaccumulations.

We assess the performance tradeoffs in terms of both memory and AD-specific time (aggregated time spent on recording, management, and evaluation), which we plot for selected configurations and their varying degrees of parallelism in the right part of \cref{fig:single_socket_memory}. We consider the classical MPI build with linear identifier management and the fully optimized hybrid build. Clearly, the former dominates the latter in terms of both memory and runtime for the Onera M6 test case. While the hybrid capabilities allow trading memory for runtime at higher degrees of parallelism in the NACA 0012 test case, the MPI build with linear identifier management at 8-fold parallelism still dominates higher degrees of hybrid parallelism. This suggests that, without further improvements, the classical MPI parallelism is the preferred choice for single-socket discrete adjoint computations in SU2. The same recommendation was already made regarding the choice of parallelism for primal single-socket computations in SU2 \cite{GomesEP2021}. In the right part of \cref{fig:single_socket_memory}, we also consider the MPI build with reuse identifier management and only hybrid AD compatible preaccumulations, which is the MPI build that is most similar to the hybrid build in terms of AD and admits a fair assessment of the hybrid AD performance. Starting already at small degrees of parallelism, we see that hybrid AD can serve as a tool to balance memory consumption and runtime.

The results also broaden the perspective on the hybrid parallel discrete adjoint performance. For instance, we expect the notable difference in the memory scaling (left part of \cref{fig:single_socket_memory}) to continue for larger numbers of cores, which could make hybrid parallel discrete adjoints attractive on CPUs with higher core counts. Any principal improvement of hybrid parallel AD such as reduction of the memory offset or improvements of the evaluation time could bridge the small margin between the classical MPI build and the fully optimized hybrid build for the NACA 0012 test case (top right part of \cref{fig:single_socket_memory}). Finally, as the memory overhead in the Onera M6 test case (bottom left part of \cref{fig:single_socket_memory}) is mostly due to disabling preaccumulations, we identify them as a promising target for further improvements of the hybrid parallel discrete adjoint performance.

\subsection{Loop parallelization strategies}
\label{sec:loop_parallelization}

\begin{figure}
    \begin{center}
        \includegraphics[width=\textwidth]{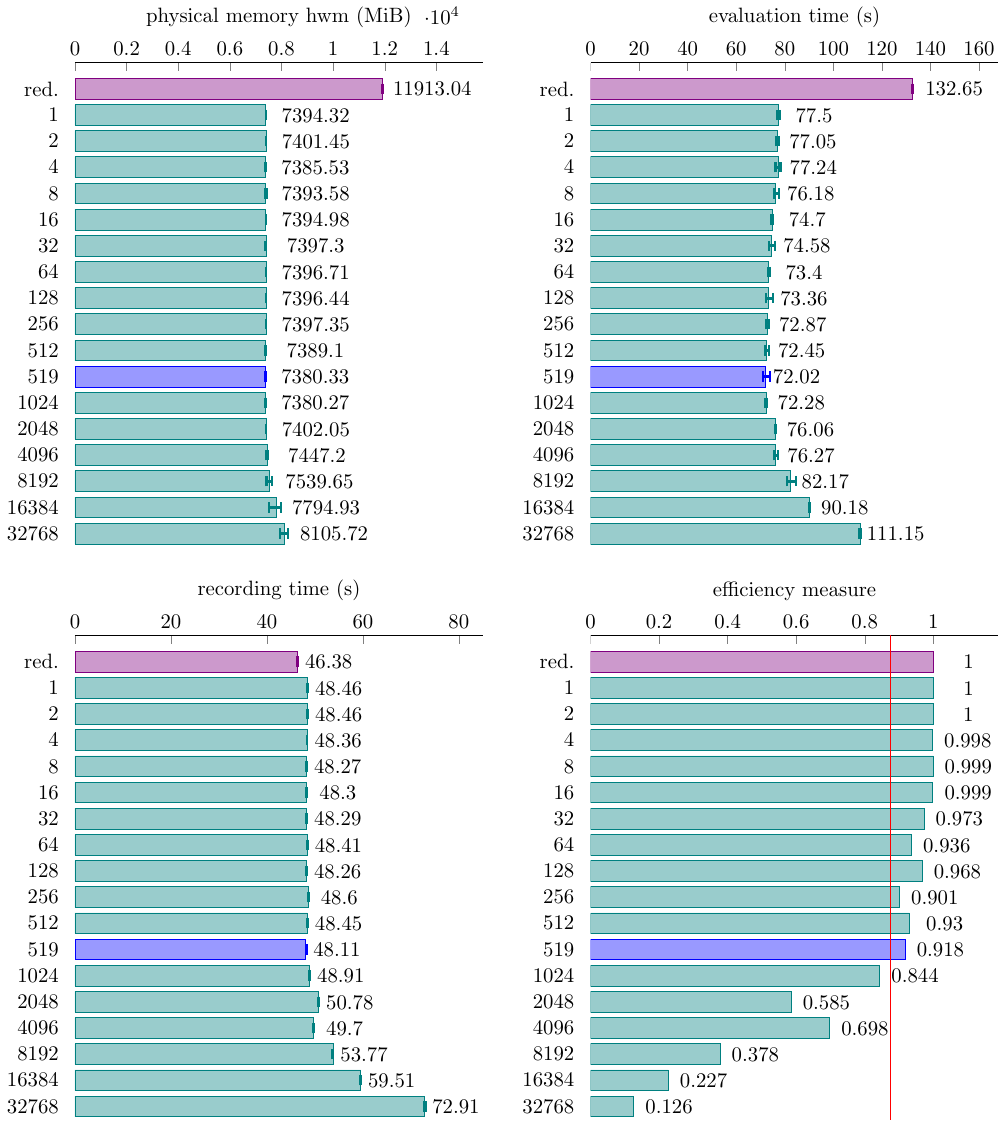}
    \end{center}
    \caption{Performance of the Onera M6 test case. We compare reductions (violet) to edge coloring with various edge color group sizes
    (teal and blue). Our adaptive algorithm chooses an edge color group size of 519 (blue). The performance plots show physical memory high-water marks (top left), evaluation time (top right), and recording time (bottom left) together with the variation across multiple runs. The efficiency measure (bottom right) quantifies the imbalance resulting from edge color groups, with an efficiency threshold of 0.875 (red).}
\label{fig:oneram6_performance}
\end{figure}

We conduct a performance study to support the design decisions regarding loop parallelization for discrete adjoints as outlined in \cref{sec:performance_optimizations}. We study the performance of the Onera M6 test case described in \cref{sec:single_socket} with both the reduction strategy and edge coloring for loop parallelization, where we vary the edge color group size in the latter. We also assess the performance of the adaptive choice of the edge color group size that we motivated in \cref{sec:performance_optimizations}. The results are displayed in \cref{fig:oneram6_performance}.

The memory comparison (top left) and the evaluation timings (top right) clearly show that edge coloring is important for the discrete adjoint performance. Even small edge color group sizes result in reduced memory usage and notably improved evaluation performance. As long as the edge color group size does not become too large, the specific edge color group size is of secondary importance. The reduction strategy requires more overall operations and more memory also without discrete adjoints, so that with discrete adjoints, both the increased number of operations and the non-exclusive read-access that prevents preaccumulation result in larger tapes, hence the increase memory usage. The non-exclusive read access also prevents shared reading optimizations, thus results in evaluations with more atomic updates. This and the larger tapes explain the increased evaluation times with reductions. The recording (bottom left), on the other hand, becomes faster with reductions despite more overall operations due to saving the effort of preaccumulation. However, the overall impact on the recording performance is small. The efficiency plot (bottom right) shows that the assumption of a monotonic relationship between edge color group size and efficiency is justified in the sense that it can serve as a basis for a heuristic selection of a large, admissible edge color group size. With the efficiency threshold of 0.875 that already decides between edge coloring and reductions in the primal solvers, the adaptive algorithm performs reasonably well at selecting an admissible edge color group size with minimal runtime, in particular minimal evaluation time.

\subsection{Scalability}
\label{sec:large_scale}

\begin{figure}
\begin{center}
\includegraphics[width=0.49\textwidth]{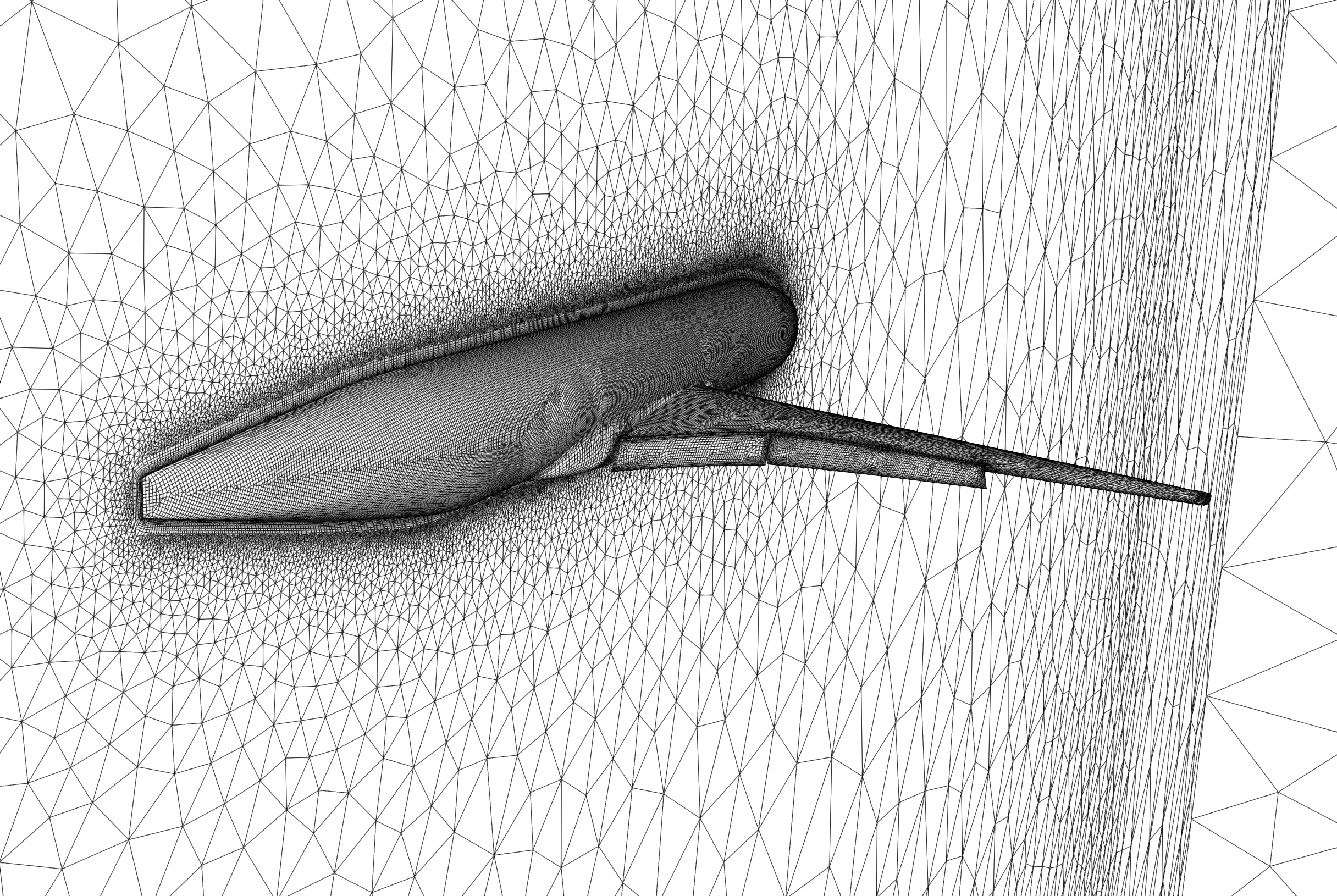}
\includegraphics[width=0.49\textwidth]{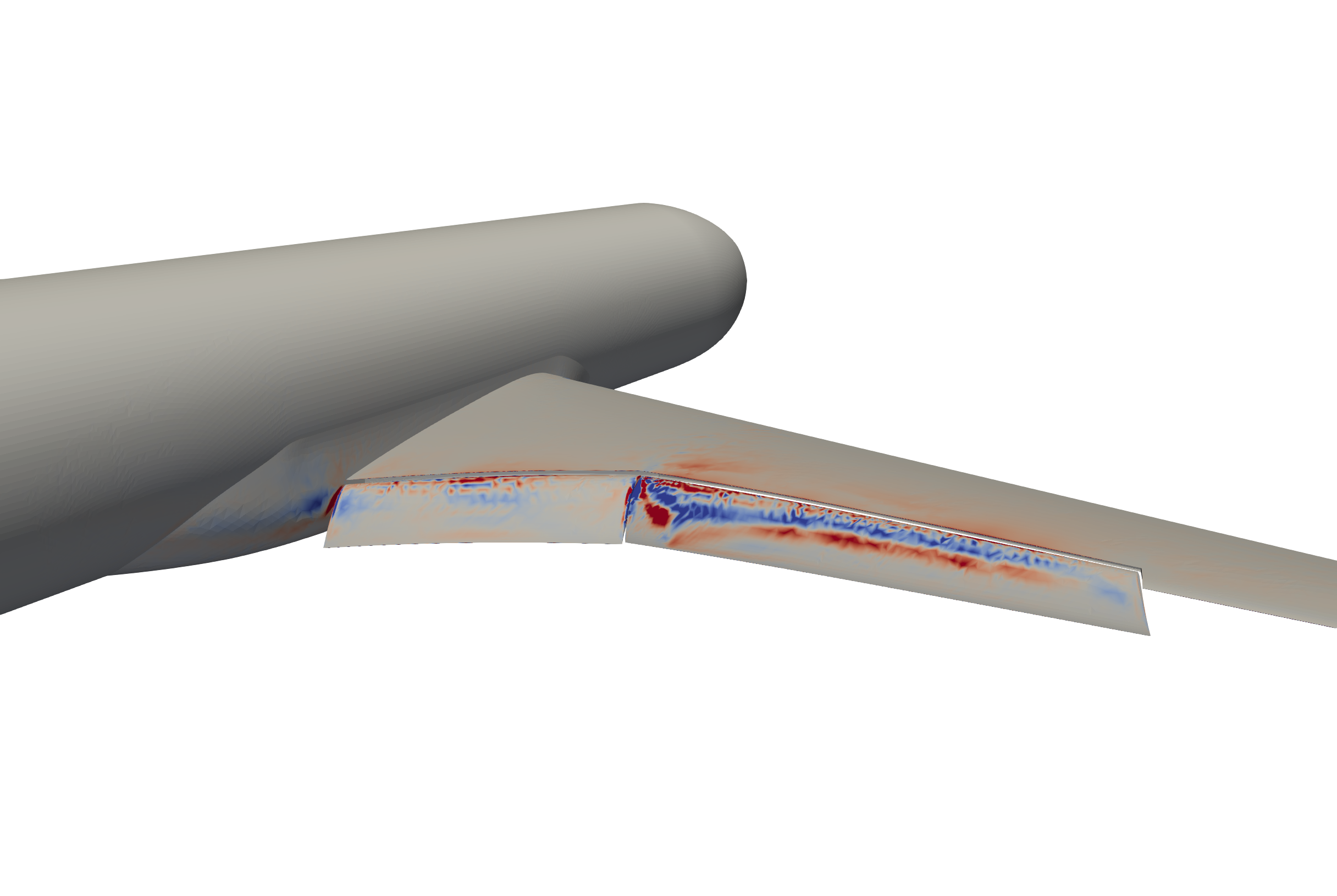}
\end{center}
\caption{HL-CRM mesh for scaling studies (left) and HL-CRM surface sensitivity (right). The latter is obtained from the derivative of the aerodynamic lift with respect to all mesh coordinates in an additional projection step, treating only surface nodes as design variables.}
\label{fig:scaling_mesh}
\end{figure}
To assess the scalability of the AD-specific performance, we use an official test case of the Third AIAA CFD High Lift Prediction Workshop involving NASA's High Lift Common Research Model (HL-CRM) \cite{RumseySS2019}, with $8^\circ$ AoA at Mach 0.2, $\text{Re}=3.26\cdot 10^6$ and a Reynolds length of 275.8\,inches. Like in the Onera M6 test case, the dynamic viscosity follows Sutherland's law with parameters for air. This test case was also used to assess SU2's hybrid parallel primal performance in \cite{GomesEP2021}. We work with the coarse version of an unstructured mixed-element HL-CRM mesh that consists of approximately 18 million cells and 8.3 million nodes, displayed together with the surface sensitivity in \cref{fig:scaling_mesh}. We start with the converged primal solution and solve \cref{eq:adjoint_linear_system} with 1000 GMRES iterations, restarting the Krylov solver every 100th iteration. Afterwards, we evaluate \eqref{eq:adjoint_sensitivity} once.

We consider the classical MPI build with linear management, the MPI build with reuse management, and the MPI build with reuse management and only hybrid AD compatible preaccumulations. We compare all of the aforementioned to the fully optimized hybrid build. Due to the memory requirements of this test case, we start with 8 nodes (192 cores) and scale up to 32 nodes (768 cores). With MPI builds, we use one MPI process per core, bound to the respective socket. We execute the hybrid build with one MPI process per socket, bound to the socket, and 12 OpenMP threads per MPI process (one per core).

\begin{figure}[t]
    \includegraphics[width=\textwidth]{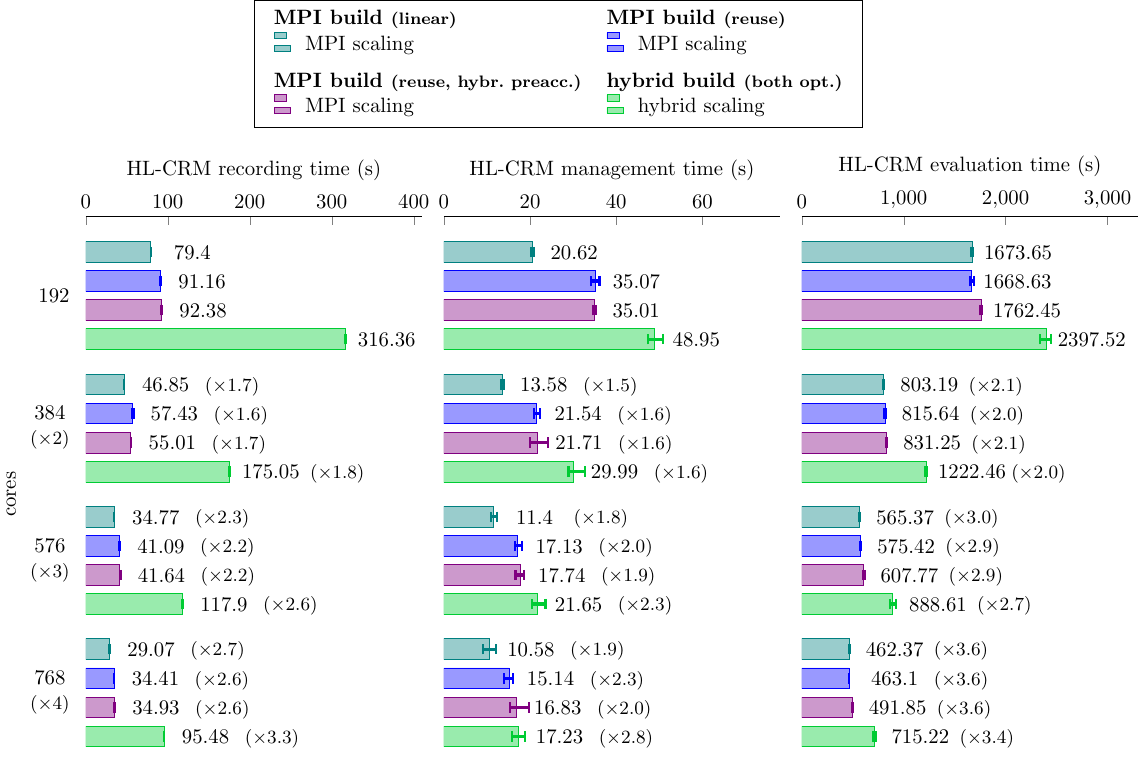}
    \caption{AD-specific performance of the HL-CRM test case. Recording performance (left), management performance (middle), and evaluation performance (right). Parallel timings for various build configurations, ranging from 192-fold to 768-fold parallelism. Error bars indicate the variation across runs. Speedups are relative to the respective timings at 192-fold parallelism.}
    \label{fig:hlcrm_performance}
\end{figure}
\cref{fig:hlcrm_performance} showcases the recording, management, and evaluation performance for the HL-CRM test case for the build configurations with various degrees of parallelism. The trends agree with the previous observations in \cref{sec:single_socket}. Considering the recording performance, the relative difference between MPI builds and the hybrid build stands out in this plot. However, this difference is actually similar to or smaller than the corresponding differences in the Onera M6 test case (comparing 12-fold MPI parallelism with 12-fold OpenMP parallelism). Even though the recording is not the major part of the AD-specific runtime, it could be worthwhile to identify and, if possible, improve its parts that do not scale with OpenMP. We see that sufficiently large degrees of parallelism close the gap in the management performance between the hybrid build and the most similar non-hybrid build.

\begin{figure}[t]
    \begin{center}
        \includegraphics[width=0.8\textwidth]{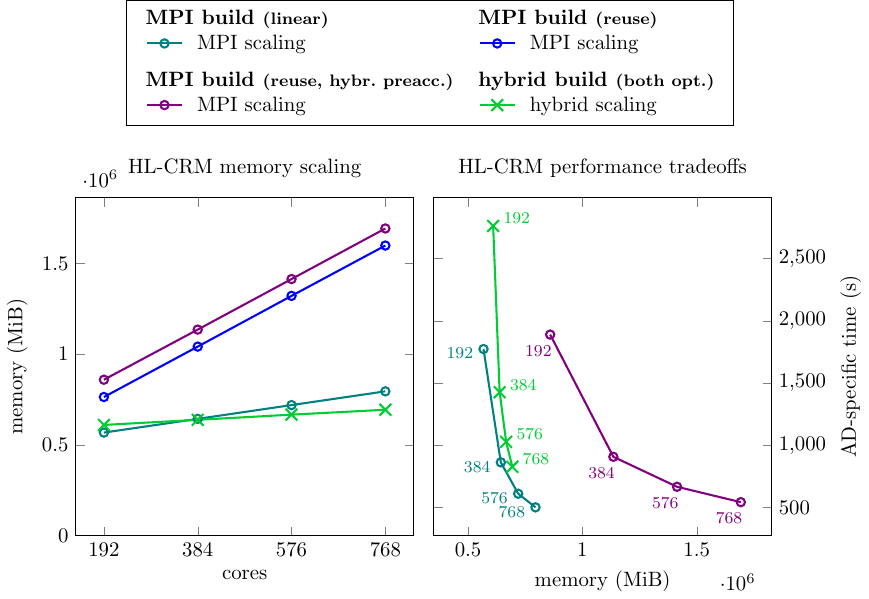}
    \end{center}
    \caption{HL-CRM memory high-water marks for the four builds and varying degree of parallelism (left), and joint memory consumption and evaluation time for selected configurations with varying degrees of parallelism (right). Plots display joint memory usage across all involved nodes.}
    \label{fig:hlcrm_memory}
\end{figure}
The left part of \cref{fig:hlcrm_memory} displays the memory consumption of the HL-CRM test case with the different build configurations and depending on the degree of parallelism. The plot clearly shows that reuse identifier management does not only introduce an offset in the memory consumption (as observed in the previous test cases) but also affects the scaling of the memory usage. More MPI processes lead to a larger overall number of halo cells and a larger overall number of copy operations for the associated MPI buffers. Unlike recordings with linear identifier management, approaches with reuse identifier management have to record copy operations on the tape, which could explain why their memory usage grows faster with an increasing number of MPI processes. Even though hybrid builds rely on a reuse management scheme as well, introducing 12-fold OpenMP parallelism per MPI process is still sufficient to achieve better memory scaling than the MPI build with linear identifier management. The crossover point lies at approximately 384-fold parallelism for this test case. If we compare the memory consumption of the hybrid configuration with the most similar non-hybrid configuration (MPI, reuse, hybr.~preacc.), we see that the hybrid AD approach can save substantial amounts of memory, there is a reduction by a factor of approximately $\times2.4$ at 768-fold parallelism. The plot also shows the additional memory consumption due to disabling preaccumulations, clearly visible as the offset between the curves for MPI builds with reuse management and enabled and disabled preaccumulations, respectively. Re-enabling preaccumulations in the hybrid build could improve its memory consumption up to the same offset. It would shift the crossover point to notably smaller degrees of parallelism and approximately double the memory savings at 768-fold parallelism due to moving from the MPI build with linear management and MPI parallel execution to a hybrid build with hybrid parallel execution in this test case. We expect further improvements on CPUs with larger numbers of cores that admit higher degrees of OpenMP parallelism per MPI process.

The right part of \cref{fig:hlcrm_memory} summarizes the performance in terms of both memory consumption and AD-specific runtime. If we compare the hybrid configuration with the most similar non-hybrid configuration in terms of AD, we see that hybrid AD can be used to trade memory for runtime also in this large test case and correspondingly higher degrees of parallelism. While there are similar tradeoffs between the hybrid configuration and the MPI configuration with linear management, the runtime price for memory reductions is relatively high in this comparison. The plot suggests that hybrid parallel configurations have MPI parallel counterparts with lower degrees of parallelism that consume similar amounts of memory but are faster. The latter are the preferable choice as long as the correspondingly smaller number of nodes offers sufficient memory, at least until further improvements of the hybrid AD performance are in place.

\section{Conclusion and future work}
\label{sec:conclusion}

Building on recently added support for OpenMP-parallel primal computations and previous support for MPI-parallel discrete adjoints, we extended SU2 by capabilities to parallelize discrete adjoint computations by both MPI and OpenMP. As a key step, we applied OpDiLib, an add-on for operator overloading AD tools that enables differentiation of OpenMP-parallel codes, and thereby demonstrated its applicability to a large code base. To properly integrate this new feature, we revisited all parts of SU2's advanced AD workflow. We identified parts of the SU2 code that need changes, discussed the underlying issues, and showcased how the code can be adapted accordingly. As some of the changes involved tradeoffs in terms of code maintainability, we proposed suitable automatic tests to facilitate future developments. We thereby established sustainable support for hybrid parallel discrete adjoints in SU2.

We conducted detailed performance studies to understand the performance characteristics of the hybrid AD approach both with respect to memory usage and AD-specific runtime. We identified the strategy for identifier management, preaccumulations, and atomic updates on adjoint variables as key differences to the previous MPI-parallel differentiation strategy in SU2 and quantified their impact on the performance in various test cases. Throughout all tests, we observed that OpenMP consistently leads to better memory scaling. As we compared the hybrid AD approach to a non-hybrid AD approach with equivalent identifier management and preaccumulation, we could demonstrate a significant reduction in memory usage for highly parallel setups, which is an important insight given the substantial memory cost of tape-based reverse AD. However, we also observed notable runtime overheads of hybrid AD, especially compared to an MPI approach with linear management and all preaccumulations, so that hybrid AD could only serve as a tool to trade memory for runtime in our test cases. With the shared reading optimization and optimized adjoint vector management, we already applied two initial improvements of hybrid AD and showcased their effectiveness in our performance tests. However, these only mark the first steps towards efficient hybrid parallel discrete adjoints in SU2. Right now, hybrid parallel discrete adjoints could be useful for problems that exceed available memory capacities, especially on CPUs where large numbers of OpenMP threads per MPI process can be used efficiently, and for large numbers of discrete adjoint iterations.

Our observations suggest various perspectives for future improvement. Within SU2, it could be worthwhile to investigate and, if possible, improve the scalability of the recordings, even though it is not the dominant part of the AD-specific runtime and does not increase with the number of discrete adjoint iterations. The evaluation time and its scalability could be improved by eliminating more atomic updates, either by identifying further exclusive read access or by revising the data access patterns to reduce non-exclusive read access. Although the scaling factors for evaluations are competitive already, they would need to be larger (or break down later) than the MPI scaling factors so that hybrid AD outperforms MPI-parallel AD not only in terms of memory but also in terms of runtime for high degrees of parallelism. Our studies suggest that the parallel reuse identifier management that is currently in place for the hybrid AD approach might not be the best choice for SU2. It could be beneficial to move to a parallel single-use scheme that supports optimized copies, with the additional benefit that it would readily support the multizone taping approach, but with the drawback that copies behave like reference in terms of exclusive read access. We plan to evaluate these improvements in further studies. As we disabled simultaneous preaccumulations with shared inputs, we expect that approaches that re-enable them lead to reduced memory usage, however, likely accompanied by new memory and/or runtime overheads in the recording phase. This issue is addressed in detail in follow-up research in \cite{BluehdornG2024}, where we propose and evaluate approaches that re-enable these preaccumulations.

While the focus of this work is on single-zone problems and the RANS solver, the insights and techniques extend readily to other solvers within SU2 and to applications of OpDiLib to other software packages and AD workflows. As we fully integrated support for hybrid parallel discrete adjoints in SU2, SU2 can serve as a testbed for any future improvement of hybrid parallel AD, and, in turn, directly benefit from improved performance.

\newpage

\section*{Acknowledgements}

Johannes Bl{\"u}hdorn, Max Aehle, and Nicolas R.~Gauger gratefully acknowledge funding from the German National High Performance Computing (NHR) association for the Center NHR South-West.

\bibliography{literature}

\end{document}